\documentclass{aa}
\usepackage{graphicx}
\usepackage{wasysym}
\usepackage{amssymb}
\usepackage{amsmath}

\usepackage{graphicx} % for plots/pictures
\usepackage[latin1]{inputenc} % Encodage du texte en entr<E9>e (accents, etc...)
\titlerunning{About the contribution in Venus rotation rate variation}
\authorrunning{Cottereau et al.}

\begin{document}

\title{About the various contributions in Venus rotation rate and LOD}

\author{L.~Cottereau \inst{1}, N.~Rambaux \inst{2, 3}, S.~Lebonnois \inst{4}, J.~Souchay\inst{1} }

%\offprints{J. Souchay \email{Jean.Souchay@obspm.fr}}
\institute{\inst{1} Observatoire de Paris, Syst\`emes de R\'ef\'erence Temps Espace (SYRTE), UMR 8630, CNRS, Paris, France\\ \quad (Laure.Cottereau, Jean.Souchay@obspm.fr) \\ \inst{2}Universit\'e Pierre et Marie Curie, Paris 6, \inst{3}IMCCE, UMR 8028, CNRS Observatoire de Paris, France (Nicolas.Rambaux@imcce.fr)\\ \inst{4}Laboratoire de m\'et\'eorologie dynamique, UMR 8539, IPSL, UPMC, CNRS, France (Sebastien.Lebonnois@lmd.jussieu.fr)}

\date{}
\abstract
% context heading (optional)
{Thanks to the Venus Express Mission, new data on the properties of Venus could be obtained in particular concerning its rotation. }
% aims heading (mandatory)
{In view of these upcoming results, the purpose of this paper is to determine and compare the major physical processes influencing the rotation of Venus, and more particularly the angular rotation rate.}
% methods heading (mandatory)
{Applying models already used for the Earth, the effect of the triaxiality of a rigid Venus on its period of rotation are computed. Then the variations of  Venus rotation caused by the elasticity, the atmosphere and the core of the planet are evaluated.}
% results heading (mandatory)
{Although the largest irregularities of the rotation rate of the Earth at short time scales are caused by its atmosphere and elastic deformations, we show that the Venus ones are dominated by the tidal torque exerted by the Sun on its solid body. Indeed, as Venus has a slow rotation, these effects have a large amplitude of $2$ minutes of time (mn). These variations of the rotation rate are larger than the one induced by atmospheric wind variations that can reach $25-50$ seconds of time (s), depending on the simulation used. The variations due to the core effects which vary with its size between $3$ and $20$s are smaller. Compared to these effects, the influence of the elastic deformation cause by the zonal tidal potential is negligible.}
% conclusions heading (optional), leave it empty if necessary 
{As the variations of the rotation of Venus reported here are of the order $3$mn peak to peak, they should influence past, present and future observations providing  further constraints on the planet internal structure and atmosphere.}

\keywords{Venus rotation}

\maketitle
\section{Introduction}

The study of the irregularities of the rotation of a planet provides astronomers and geophysicists with physical contraints on the models describing this planet and allows a better understanding of its global properties.  Although Venus is the planet sharing the most similarities with the Earth in terms of size and density, some of its characteristics are poorly understood like its atmospheric winds and its superrotation. With its thick atmosphere, the determination of the period of rotation of Venus has been  and still is a very challenging task. Today thanks to Venus Express (the first spacecraft orbiting Venus since the Magellan Mission in 1994) and Earth-based radar measurements, the period of rotation of Venus can be revised and its variations evaluated. So here for the first time a complete theoritical study of the variations of the rotation of Venus on a short time scale is presented as well as its implications concerning the observations.

Many studies on the rotation of Venus have already been made. Several authors studied this rotation on a long time scale to understand why Venus spin is retrograde and why its spin axis has a small obliquity (Goldstein 1964; Carpenter 1964; Goldreich and Peale 1970; Lago and Cazenave 1979; Dobrovoskis 1980; Yoder 1995; Correia and Laskar 2001, 2003), others studied the possible resonance between the Earth and Venus (Gold and Soter, 1979; Bills 2005; Bazs\'o et al., 2010). But few studies have been made comparing the major physical processes influencing the rotation of Venus at short time scale. Karatekin et al. (2010) have shown that the variations of the rotation of Venus due to the atmosphere should be approximatively 10 seconds for the $117$ days time span characterizing the planet's solar day, but they neglected the main effect of the impact of gravitational torques on the solid body of the planet.
Here, applying different models already used for the Earth, the irregularities of the rotation of Venus are computed. After clarifying the link between the length of day ($LOD$) and the rotation rate of the planet, the effects of the triaxiality of a rigid Venus on its period of rotation are evaluated. This model is based on Kinoshita's theory (1977) and uses the Andoyer variables (Andoyer, 1923). As Venus has a very slow rotation (-243.020 d), these effects have a large amplitude (2 mn peak to peak) and could be observable as this is shown in Section \ref{solide}.

As we know, the variation of the period of rotation of the Earth of the order of $0.6$ milliseconds (ms) for the seasonal component, are principaly due to the tidal zonal potential, the atmosphere and the oceans (Munk and MacDonald, 1960; Lambeck, 1980; Barnes et al., 1983). Hereafter quantifying the effect of the zonal component of the solar tides on Venus (Section \ref{zonal}), the effects of its atmosphere (Section~\ref{atm}) and the impact of its internal structure (Section \ref{noyau}) are presented.

To conclude, a comparison of the properties and amplitudes of these various processes is analysed (Section \ref{discussion}). Their observability and their implications concerning the properties of Venus are also discussed. We find that, on the opposite of what is often assumed in the literature, the atmosphere should not be the most important effect leading to variations of the rotation rate of Venus. Indeed, the tidal torque exerted by the Sun on the solid body of the planet should be larger by more than a factor of two.

\section{Rotation rate and LOD definition}\label{def}

A day is defined as the time between two consecutive crossings of a reference meridian by a reference point or body. While these periods vary with respect to time, we can use their average values for the purpose. For the Earth, the solar day (24 h) and the sidereal day (23.56 h)  are defined respectively when the Sun and a star are taken as reference. They are very close because the Earth's period of rotation is far smaller than its period of revolution around the Sun. By contrast, as Venus has a slow rotation, its solar and sideral days are quite different. Indeed, the canonical value taken for its rotational period (sideral day) is $243.02$d (Konopliv et al, 1999), whereas its solar day varies around $117$d.  In the following, to be consistent with the physical inputs in the global circulation model $(GCM)$ simulations, we will fix the value of the solar venusian day at $117$d.\\
There are mainly two methods to measure the rotation period, and its variations. The radar Doppler measurements give direct access to the instantaneous rotation rate $\omega (t)$ that we can translate in an instantaneous period of rotation $LOD(t)$ by 
\begin{equation}
LOD(t)=\frac{2\pi}{\omega(t)}.
\end{equation}
Because we will be interested in variation around a mean value, we further define $\Delta \omega (t)=\omega (t)-\overline{\omega}$ and $\Delta LOD (t)=LOD (t)-\overline{LOD}$ where $\overline{LOD}=243.02$d and $\overline{\omega}=\frac{2\pi}{\overline{LOD}}$. At first order  $\frac{\Delta \omega}{\overline{\omega}}(t)=-\frac{\Delta LOD}{\overline{LOD}}(t)$.
Thanks to infrared images of the surface of the planet, we can also measure the longitude of a reference point $\phi_{1}$ with respect to a given reference system at several epochs where $\dot{\phi_{1}}=\omega(t)$. As we will discuss in Section \ref{discussion}, as $\omega(t)$ has periodic variations, fitting $\phi_{1}$ by a linear function could lead to a wrong estimation of $\bar{\omega}$.

\section{Effect of the solid potential on the rotation rate} \label{solide}

A first and simplified model to describe the variation of the rotation of Venus is to consider that the atmosphere, mantle and core of Venus are rotating  as a unique solid body. This model enables us to use Kinoshita's theory (1977) with the Andoyer Variables (1923) as those describing the rotation of the Earth. The rotation of Venus is then described by three action variables ($G,L,H$) and their conjugate variables ($g,l,h$). $G$ represents the amplitude of the angular momentum and  $L,H$ respectively its projections on the figure axis and on the inertial axis (axis of the reference plane) such as
\begin{equation}
L=G\cos J \  \mathtt{and} \ H=G \cos I
\end{equation}
where $I$ and $J$ correspond respectively to the angles between the angular momentum axis and the inertial axis and between the angular momentum axis and the figure axis. Here the figure axis and the inertial axis coincide respectively with the axis of the largest moment of inertia and the axis of the orbit of Venus at J2000.0. The angle $J$ is yet unknown. As shown in appendix A, for plausible value of this angle, its impact on the rotation is weak with respect to the other effects taken into account in this article. For sake of clarity, $J$ is set to $0$ in the following.
In this coordinate system, the Hamiltonian related to the rotational motion of Venus is (Cottereau and Souchay, 2009):
\begin{equation}
K=F_{0}+ E+E'+U.
\end{equation}
$F_{0}$ is the Hamiltonian for the free rotational motion defined by
\begin{equation} \label{ham}
F_{0}= \frac{1}{2}(\frac{\sin ^{2} l}{A}+\frac{\cos ^{2} l}{B})(G^{2}-L^{2})+\frac{1}{2}\frac{L^{2}}{C},
\end{equation}
where $A, B, C$ are the principal moments of inertia of Venus. $E, E'$ are respectively the components related to the motion of the orbit of Venus which is caused by planetary perturbations (Kinoshita, 1977) and to the choice of the "departure point" as reference point (Cottereau and Souchay, 2009). $U$ is the disturbing potential of the Sun, considered as a point mass, and is given at first order by :
\begin{eqnarray}\label{eq1}
U&&=\frac{\mathcal{G} M'}{r^3}\Big[[\frac{2C-A-B}{2}]P_{2}(\sin \delta )\nonumber\\&&+[\frac{A-B}{4}]P_2^{2} (\sin \delta) \cos 2\alpha\Big],
\end{eqnarray}
where $\mathcal{G}$ is the gravitational constant, $M'$ is the mass of de Sun, $r$ is the distance between its barycenter and the barycenter of Venus. $\alpha$ and $\delta$ are the planetocentric longitude and latitude of the Sun with respect to the mean equator of Venus and a meridian origin (not to be confused with the usual equatorial coordinates defined with respect to the Earth). The $P_{n}^m$ are the classical Legendre functions given by:
\begin{equation}\label{legendre}
P_{n}^m (x)= \frac{(-1)^m(1-x^2)^{\frac{m}{2}}}{2^n n!}\frac{d^{n+m} (x^2-1)^n}{dx^{n+m}}.
\end{equation}
The hamiltonian equations are :
\begin{eqnarray}\label{ecano1}
\frac{\mathtt{d}}{\mathtt{d}t}(L,G,H)= -\frac{\partial K}{\partial(l,g,h)}
\end{eqnarray}
\begin{eqnarray}\label{ecano2}
\frac{\mathtt{d}}{\mathtt{d}t}(l,g,h)= \frac{\partial K}{\partial(L,G,H)}. 
\end{eqnarray}
As the components $\omega_{1}$ and $\omega_{2}$ of the rotation of Venus are supposed at first approximation to be negligible with respect to the component $\omega_{3}$ along the figure axis, this yields :
\begin{eqnarray}\label{ecano2}
G=\sqrt{(A\omega_{1})^2+(B\omega_{2})^2+(C\omega_{3})^2}\approx C\omega_{3}
\end{eqnarray}
and
\begin{eqnarray}\label{ecano2}
\frac{\mathtt{d}}{\mathtt{d}t}(C\omega_{3})\approx \frac{\mathtt{d}}{\mathtt{d}t}(C\omega)\approx -\frac{\partial K}{\partial g }. 
\end{eqnarray}
Splitting $\omega$ into its mean value $\overline{\omega}$ and a variation $\Delta \omega$, $\omega$ is given by :
\begin{equation}\label{eqO}
\frac{\mathtt{d}}{\mathtt{d}t}(C\Delta \omega)= -\frac{\partial K}{\partial g}. 
\end{equation}
Notice here that $F_{0}$ and $E+E'$ do not contain $g$, so that the variations of the rotation of solid Venus are only caused by the tidal torque of the Sun included in the disturbing potential $U$.
To make explicit the dependence on the variable $g$, we express $U$ as a function of the longitude $\lambda$ and the latitude $\beta$ of the Sun with respect to the orbit of Venus at the date $t$ with the transformations described by Kinoshita (1977) and based on the Jacobi polynomials such as :
\begin{eqnarray}\label{appl}
\frac{d(C\Delta \omega)}{dt}&&=-\frac{\partial }{\partial g}\frac{\mathcal{G} M'}{r^3}\Bigg[\frac{2C-A-B}{2}\Big(-\frac{1}{4}(3\cos ^2 I-1)\nonumber\\&&-\frac{3}{4}\sin^2 I \cos 2 (\lambda-h)\Big)+\frac{A-B}{4}\nonumber \\&& \bigg[ \frac{3}{2} \sin^2I \cos(2l+2g)+\sum_{\epsilon=\pm 1}\frac{3}{4}(1+\epsilon \cos I)^2\nonumber \\&& \cos 2(\lambda-h-\epsilon l-\epsilon g)\bigg]\Bigg],
\end{eqnarray}
where $\beta$ is, by definition, equal to 0 in this case. The first component at the right hand-side of Eq.(\ref{appl}), does not depend on the variable $g$, so that this yields:
\begin{eqnarray}\label{eq3}
\Delta \omega&&=-\frac{1}{C}\int \frac{\partial}{\partial g}\Bigg(n^2\Big(\frac{a}{r}\Big)^3 \frac{A-B}{4}\bigg[ \frac{3}{2} \sin^2I \cos(2l+2g)\nonumber \\&&+\sum_{\epsilon=\pm 1}\frac{3}{4}(1+\epsilon \cos I)^2 \nonumber \\&& \cos 2(\lambda-h-\epsilon l-\epsilon g)\bigg]\Bigg)\mathtt{d}t.
\end{eqnarray}
where $\frac{\mathcal{G}M'}{r^3}$ has been replaced by $n^2 \cdot \frac{a^3}{r^3}$, $a$ and $n$ being respectively the semi-major axis and the mean motion of Venus defined by the third Kepler law
\begin{equation}
n^2a^3=\mathcal{G}M'.
\end{equation}
Differentiating with respect to $g$ in the Eq.(\ref{eq3}), $\Delta{\omega}$ is given by :
\begin{eqnarray}\label{eq4}
\Delta \omega&&=-\int 3\frac{A-B}{4C}\,n^2\Big(\frac{a}{r}\Big)^3\bigg[- \sin^2I \sin(2l+2g)\nonumber \\&& +\sum_{\epsilon=\pm 1}\frac{1}{2}\epsilon(1+\epsilon \cos I)^2\nonumber \\&&\sin 2(\lambda-h-\epsilon l-\epsilon g)\bigg]\mathtt{d}t.
\end{eqnarray}
Notice that we can write $l+g \approx \phi$ (Kinoshita, 1977) where $\phi$ is the angle between the axis which coincides with the smallest moment of inertia and a reference point arbitrarly chosen as the "departure point" on the orbit of Venus at the date $t$ (Cottereau and Souchay, 2009). The position of this axis is given with respect to the origin meridian itself defined as the central peak in the crater Adriadne (Konopliv et al, 1999; Davies et al., 1992). In fact $\phi$ is the angle of proper rotation of the planet ($\dot{\phi}\approx\omega$). In the following, $\phi$ will be used instead of $l+g$. To solve analytically Eq.(\ref{eq4}), the developments of $(\frac{a}{r})^3 \sin 2\phi$ and $(\frac{a}{r})^3\sin 2(\lambda-h\pm \phi)$ are needed as a function of time through the variables $M$ and $L_{S}$ (respectively the mean anomaly and the mean longitude of Venus) taking the eccentricity as a small parameter (Kinoshita, 1977). In a first approach, the orbit of Venus can be considered as circular (i.e : $e$=0, instead of : $e$=0.0068). This yields :
\begin{eqnarray}\label{eq5}
\Delta \omega&&=3 \frac{B-A}{4C}\,n^2\bigg[ \sin^2I \,\cos(2\phi)\,\frac{1}{2\dot{\phi}}\nonumber \\&&
-\frac{1}{2}(1+\cos I)^2 \cos 2(L_{s}-\phi)\frac{1}{2\dot{L_{s}}-2\dot{\phi}} \nonumber \\&&+\frac{1}{2}(1-\cos I)^2\cos 2(L_{s}+\phi)\,\frac{1}{2\dot{L_{s}}+2\dot{\phi}}\bigg].
\end{eqnarray}
Notice that this equation is similar to the equation given by Woolard (1953) in his theory of the Earth rotation, using a different formalism based on classical Euler angles. 
\begin{table}[!h]
\begin{center}
\resizebox{1\hsize}{!}{\begin{tabular}[h]{l|c}
\hline\\
  & Venus \\
\hline \hline\\
Period of revolution &224.70d (Simon et al., 1994)\\
Obliquity & $2^{\circ}.6358$ (Cottereau and Souchay, 2009)\\
Mean period of rotation &-243.020d (Konopliv et al., 1999) \\
Triaxiality : $\frac{A-B}{4C}$ &-$1.647941~10^{-6}$ (Konopliv et al., 1999; Yoder, 1995)\\
$\frac{C}{M_{\mathtt{V}}R_{\mathtt{V}}^2}$ &$0.33600$ (Yoder, 1995) \\
\end{tabular}
}
\end{center}
\caption{Numerical values. $M_{\mathtt{V}}$ and $R_{\mathtt{V}}$ are respectively the mass and the radius of Venus.}
\label{values}
\end{table}

Taking the numerical values of the Table \ref{values}, the largest terms of the variation of $\frac{\Delta \omega}{\omega}$ are 
\begin{eqnarray}\label{eq7}
\frac{\Delta \omega}{\overline{\omega}}&&=2.77~10^{-6}\cos(2L_{S}-2\phi)\nonumber\\&&+6.12~10^{-9}\cos 2\phi\nonumber \\&&-1.99~ 10^{-11}\cos(2L_{S}+2\phi)
\end{eqnarray}
where $2L_{S}-2\phi, 2\phi$ and $2L_{S}+2\phi$ correspond to the leading terms with respective periods $58$ d, $121.80$ d and $1490$d. As the two last terms scale as the square of the obliquity, they are significantly smaller than the term with argument $2L_{S}-2\phi$ ($2.77~10^{-6}$). As Venus has a very slow rotation which appears in the scaling factor $3\frac{B-A}{4C}\frac{n^2}{\omega}$, the variations of the rotation rate due to the solid torque are larger than the Earth ones which correspond to amplitudes of $\frac{\Delta \omega}{\overline{\omega}}\approx 10^{-10}$.
\begin{figure}[htbp]
\center
\resizebox{0.9\hsize}{!}{\includegraphics{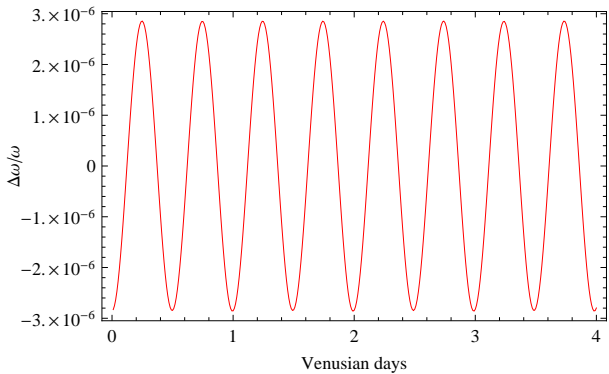}}
 \caption{Variation of $\frac{\Delta \omega}{\overline{\omega}}$ during a 4 venusian solar days time span ($468$d). One venusian solar day corresponds to $117$d.}
\label{fig1}
\end{figure}
Fig.\ref{fig1} shows the relative variations of the speed of rotation of Venus due to the solid torque exerted by the Sun during a four venusian days time span ($468$d). The orientation of the bulge of Venus at $t=0$ is given with respect to the origin meridian. The mean orbital elements are given by Simon et al. (1994). The choice of the time span will allow us to compare in the following the different effects which act on the rotation of Venus.

\begin{figure}[htbp]
\center
\resizebox{0.9\hsize}{!}{\includegraphics{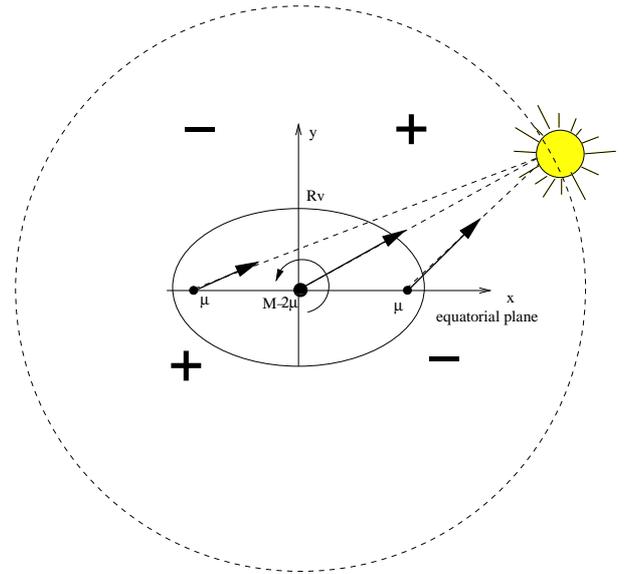}}
 \caption{Diagram of the gravitational force exerted by the Sun in the frame corotating with Venus. The $-$ and $+$ signs give the sign of the net torque in each quadrant.}
\label{bulbe}
\end{figure}

The physical meaning of the leading term with argument $2L_{S}-2\phi$ can be understood using a simple toy model. Consider a coplanar and circular orbit. At quadrupolar order, the gravitational potential created by the planet in its equatorial plane is equal to the one created by three point masses, one located at the center of the planet and of mass $M_{v}-2\mu$ and two other ones symetrically located on the surface of the planet along the axis of smallest moment of inertia and of mass $\mu$ (see Fig.\ref{bulbe}) with $\mu=\frac{B-A}{2R_{\mathtt{V}}^2}$, $R_{\mathtt{V}}$ being the planet mean radius. The torque exerted on the planet by the Sun can thus be computed using the forces exerted on these three points only. As shown in Fig.\ref{bulbe}, the sign of the resulting torque depends on the quadrant of the $x, y$ plane the Sun is located in, and thus changes four times during a solar day, whose the corresponding argument is $2L_{S}-2\phi$.

\begin{figure}[htbp]
\center
\resizebox{0.9\hsize}{!}{\includegraphics{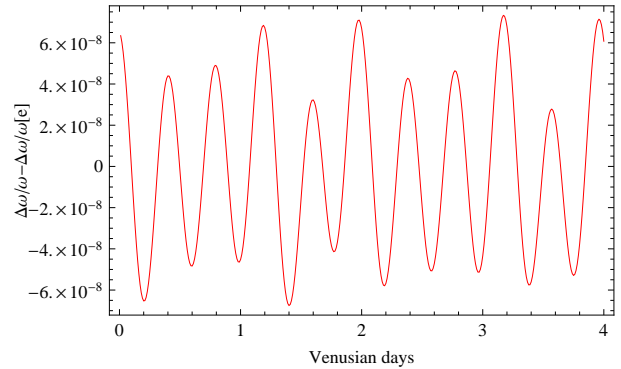}}
 \caption{Influence of the eccentricity on the variation of the speed of rotation of Venus.}
\label{fig2}
\end{figure}

To evaluate the influence of the eccentricity of Venus on its rotation rate,  we show in Fig.\ref{fig2} the residuals after substraction of $\frac{\Delta \omega}{\overline{\omega}}$ when the eccentricity is taken into account in the development of $(\frac{a}{r})^3 \sin 2\phi$ and $(\frac{a}{r})^3\sin 2(\lambda-h\pm \phi)$ with respect to the simplified expression given by the Eq.(\ref{eq7}). 

We can observe that the eccentricity of Venus acts on $\frac{\Delta \omega}{\overline{\omega}}$ with an amplitude of $\approx 10^{-8}$ for the terms $2L_{s}+M-2 \phi$ (46.34 d) and  $2L_{s}-M-2 \phi$ (78.86 d). These variations are smaller than the leading coefficient seen in Eq.(\ref{eq7}) by two orders and more important than the other coefficients with arguments $2\phi$ and $2\phi+2L_{s}$.

From the relation $\frac{\Delta \omega}{\omega}(t)=-\frac{\Delta LOD}{LOD}(t)$, the variations of the LOD(t) due to the torque of the Sun are $120$ s (i.e 2 mn or $0.0014$ d).
The implication of these results will be discussed in the following (Section \ref{discussion}). Note that these variations are larger than the uncertainties on the period of rotation measurements of Venus given by Magellan, that is to say $243.0200\pm 0.0002$d in Konopliv et al. (1999), or $243.0185\pm 0.0001$d in Davies et al. (1992) and close to the amount of their difference.

\section{Effects of the elasticity on the rotation rate : deformation due to the zonal tidal potential}\label{zonal}

\begin{figure}[htbp]
\center
\resizebox{0.9\hsize}{!}{\includegraphics{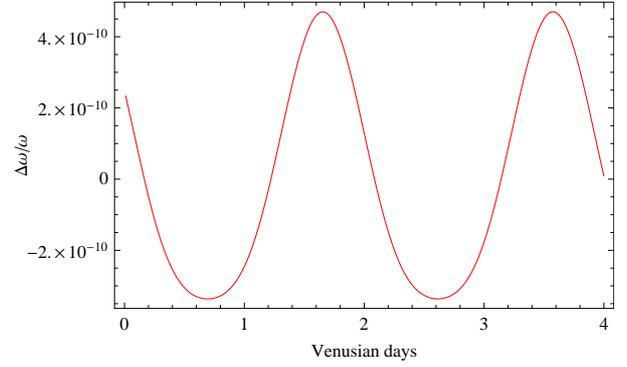}}
 \caption{$\frac{\Delta \omega}{\overline{\omega}}$ due to the zonal potential during a 4 venusian days time span ($468$d).}
\label{fig3}
\end{figure}

For the Earth, an important variation of the speed of rotation is due to the zonal potential which causes temporal variations of the moment of inertia $C$. In this section, these zonal effects are evaluated for Venus, considered as a deformable body. The zonal part of the potential exerted by the Sun on a point $M$ at the surface of the planet is given by the classical formula
\begin{eqnarray}
V_{2, \mathtt{S}}(\delta_{M},\delta)&&=\frac{9}{4} \mathcal{G}M' \frac{R_{\mathtt{V}}^2}{r^3}(\sin^2 \delta-\frac{1}{3})\nonumber\\&&
(\sin^2 \delta_{M}-\frac{1}{3}) ,
\end{eqnarray}
where $\delta$ and $\delta_{M}$ represent respectively the planetocentric latitude of the Sun (the disturbing body) and of the point $M$ and $r$ the distance between the barycenter of the Sun and the Venus one. The corresponding bulge produced has a potential (Melchior, 1978):
\begin{eqnarray}
\Delta V_{2, \mathtt{\mathtt{V}}}(\delta_{M},\delta, r')=k_{2}V_{2, \mathtt{S}}\frac{R_{\mathtt{V}}^3}{r'^3},
\end{eqnarray}
where $k_{2}$ and $R_{\mathtt{V}}$ are respectively the Love number and the radius of the planet.  Differentiating the potential produced by Venus at its surface (MacCullagh's formula), $\Delta V_{2, \mathtt{V}}$ can be expressed as a function of the principal moments of inertia
\begin{eqnarray}
\Delta V_{2, \mathtt{V}}(\delta, r')=\frac{3}{2}\frac{\mathcal{G}}{r'^3}(\mathtt{d}C-\mathtt{d}A)(\sin ^2 \delta-\frac{1}{3}),
\end{eqnarray}
where we take $\mathtt{d}A=\mathtt{d}B$ as the deformation is purely zonal.  By identification we obtain:
\begin{equation}\label{rot11}
k_{2}\frac{3}{4}\frac{M'}{M_{\mathtt{V}}}(\sin^2 \delta -\frac{1}{3})(\frac{R_{\mathtt{V}}}{r})^3=\frac{\mathtt{d}C-\mathtt{d}A}{2}\frac{1}{M_{\mathtt{V}}R_{\mathtt{V}}^2}.
\end{equation}
Because this deformation does not induce a change in the volume of the isodensity surfaces in the planet to first order, $\mathtt{d}I=\frac{1}{2}\mathtt{d}C+\mathtt{d}A=\frac{1}{2}\Delta C+\Delta A=0$ ( see Melchior, 1978, for a rigorous demonstration). The Euler's third equation is $C\omega=\mathtt{constant}$, to first order this yields $\frac{\Delta C}{C}=-\frac{\Delta \omega}{\overline{\omega}}$.
Thus the variations of the angular rate of rotation of Venus are given by :
\begin{eqnarray}\label{rots}
&&\frac{\Delta \omega}{\overline{\omega}}=-\frac{\Delta C}{C}\nonumber\\&&
=-k_{2}\Big(\frac{R_{\mathtt{V}}}{a}\Big)^3\Big(\frac{a}{r}\Big)^3 \frac{M'}{M_{\mathtt{V}}}(\sin^2 \delta -\frac{1}{3}) \frac{M_{\mathtt{V}}R_{\mathtt{V}}^2}{C}
\end{eqnarray}
where $M_{\mathtt{V}}$ is the mass of Venus.
Expressing $\sin^2 \delta_{1}$ as a function of $I$ and $\lambda$ where $I$, $\lambda$ are given in Section \ref{solide} and represent the obliquity of Venus and the true longitude of the Sun, Eq.(\ref{rots}) becomes 
\begin{eqnarray}\label{rot2}
\frac{\Delta \omega}{\overline{\omega}}&=&-k_{2}\frac{M_\mathtt{V}R_\mathtt{V}^2}{C}\frac{M'}{M_\mathtt{V}}\Big(\frac{R_\mathtt{V}}{a}\Big)^3\Big(\frac{a}{r}\Big)^3\nonumber\\&&
\Big(\frac{\sin^2 I}{2}(1-\cos 2\lambda)-\frac{1}{3}\Big).
\end{eqnarray}
Using the developments of $(\frac{a}{r})^3$ and $(\frac{a}{r})^3 \cos 2\lambda$ with respect to time from Cottereau and Souchay (2009), Fig.\ref{fig3} shows the relative variation of the speed of rotation of  Venus due to the zonal potential. The numerical values of $k_{2}=0.295\pm0.066$ and the ratio $\frac{C}{M_{\mathtt{V}}R_{\mathtt{V}}^2}=0.3360$ are taken respectively from Konopliv and Yoder (1996) and Yoder (1995). The expressions found correspond to variations of the $LOD(t)$ of $0.019$s peak to peak which is very small with respect to the contributions studied in Section \ref{solide}. So the zonal part of the potential on a non rigid Venus has a little influence on its very slow rotation, by contrast with the Earth for which these variations with annual and semi-annual component are not negligible (of the order of $10^{-3}$ s) with respect to the 1 d rotation (Yoder et al., 1981; Souchay and Folgueira, 1998). Note that the semi-annual components are particularly small for Venus because its eccentricity and obliquity are much smaller than their respective value for the Earth.

\section{Atmospheric effects on the rotation of Venus}\label{atm}

\begin{figure}[htbp]
\center
\resizebox{0.9\hsize}{!}{\includegraphics{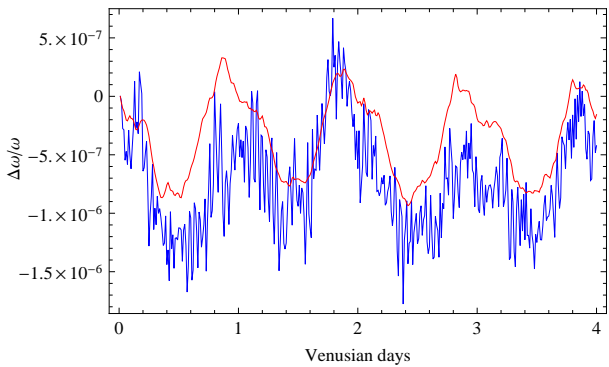}}
 \caption{$\frac{\Delta \omega}{\overline{\omega}}$ due to the atmosphere during a 4 venusian days time span for the GCM1 (red curve) and the GCM2 (blue curve).}
\label{fig4}
\end{figure}

As Venus has a denser atmosphere than the Earth, and as we know presently that the Earth atmosphere acts on its rotation in a significant manner (Lambeck, 1980), it will be interesting to study the corresponding effects on Venus.
In this section, the core and the mantle of Venus are supposed to be rigidly coupled and its atmosphere rotates at a different rate. The variations of the speed of rotation of Venus due to its atmosphere are given by :
\begin{equation}
\frac{\Delta G}{G}=\frac{\Delta \omega}{\overline{\omega}},
\end{equation}
where $G,\Delta G$ represent respectively the angular momentum of the rigid Venus and its variation due to the atmosphere $\Delta G=-\Delta G_{\mathtt{atm}}$. The angular momentum $\Delta G_{\mathtt{atm}}$ of the atmosphere can be split into two components :
\begin{itemize}
\item
The matter term $G_{\mathtt{M}}$ which is the product of $\omega$ with the inertia momentum of the atmosphere
\item
The current term $G_{\mathtt{w}}$ which is due to the wind motions with respect to the frame solidly rotating with the planet
\end{itemize}
From Lebonnois et al. (2010a) we have :
\begin{eqnarray}\label{atmosp}
G_{\mathtt{atm}}&=&(1+k_{2}')G_{\mathtt{M}}+G_{\mathtt{w}}\nonumber\\&&
=(1+k_{2}')\frac{\omega R_\mathtt{V}^4}{g}\int\int_{s} P_{s}\cos^3 \theta \mathtt{d}\theta \mathtt{d}\phi+\nonumber\\&& 
\frac{R_\mathtt{V}^3}{g}\int\int\int_{v} \cos^2 \theta v_{\theta}\mathtt{d}h \mathtt{d}\theta \mathtt{d}\phi \label{atm2}.
\end{eqnarray}
where $\theta, \phi, R_\mathtt{V}$ stand respectively for the latitude, longitude and radius of Venus,$v_{\theta}$ is the zonal wind and $k_{2}'$ is the load Love number of degree 2 (Karatekin et al., 2010). Here to determine the variations of the angular momentum, two simulations made with the global circulation model $(GCM)$ of the Laboratoire de Meteorologie Dynamique (LMD) (Lebonnois et al., 2010b) are used and compared. These simulations have been obtained with the LMD Venus General Circulation Model, using conditions similar to those presented in Lebonnois et al. (2010a), except for the boundary layer scheme. The first simulation ($GCM1$) was integrated from a zero wind state and is very close to the simulation published in Lebonnois et al. (2010a), though the winds in the deep atmosphere (0 to 40 km altitude) are slightly higher, due to the updated boundary layer scheme. The second one ($GCM2$) was integrated with initial winds in superrotation where the resulting winds in the deep atmosphere are close to observed values.  The results presented in Fig.\ref{fig4} were obtained after a $200$ days integration for $GCM1$ and $130$ days for $GCM2$. Here we calculate the seasonal variation of the angular velocity of Venus induced by the atmospheric winds to compare it with the variation discussed in section \ref{solide}. The numerical values of the radius and the principal moment of inertia are taken in Cottereau and Souchay (2009).

Figure \ref{fig4} shows the relative variation of the speed of rotation of Venus due to the atmosphere for a 4 venusian days time span (470 d) with the $GCM1$ (red curve) and $GCM2$ (blue curve). These variations have an amplitude of $1.26~10^{-6}$ peak to peak for $GCM1$ and $2.44~10^{-6}$ respectively for $GCM2$.
For the two models, the large amplitude of $\frac{\Delta \omega}{\bar{\omega}}$ of the order of $10^{-6}$ comes from the current term and depends on winds of Venus whereas the matter term acts on the rotation with an amplitude of the order of $10^{-10}$. 
 
Using the definition of the $LOD(t)$ given in Section \ref{def}, the atmospheric contributions to Venus rotational speed correspond to peak to peak variations of the $LOD(t)$ of $27$s with the GCM1 and $51$s with the GCM2. These values are consistent with the value of $LOD(t)$ of the planet of $7.9$s given by Karatekin et al. (2009, 2010),  who used the simulation presented in Lebonnois et al. (2010a). The differences are related to the amplitudes of the zonal winds in the region of maximum angular momentum (10-30 km of altitude), which vary between the different simulations. The most realistic values for these winds are obtained with the $GCM2$ simulation, where the winds are close to observational data obtained from the Venera and Pioneer Venus missions (Schubert, 1983). We compared here the results given by $GCM1$ and $GCM2$ to give an idea of the uncertainties in our present understanding of the atmospheric circulation and its modeling. In the following, only the values obtained with the model GCM2 will be compared to the other effects as it is thought to be the most realistic. Note that the atmospheric effects are nearly two times smaller than the solid effects described in Section \ref{solide}.

\section{Core effects on the rotation of Venus}\label{noyau}

The interior of Venus is probably liquid as inferred from the orbiting spacecraft data (Konopliv \& Yoder 1996). The internal properties of Venus are expected to be like the Earth with a core radius around 3120 km (Yoder 1995), but with a noteworthy difference because there is no dynamo effect on Venus (Nimmo 2002). In this Section, we investigate the impact of such a core on the rotational motion of Venus and especially on the variation of its $LOD(t)$. For that purpose, we numerically integrate the rotational motion of Venus by taking into account the inertial pressure torque. The equations governing the rotational motion of two-layer Venus are the angular momentum balance for the whole body
\begin{equation}
\frac{\mathtt{d} \vec H}{\mathtt{d}t} + \vec \omega \wedge \vec H = \Gamma
\end{equation}
and for the core	
\begin{equation}
\frac{\mathtt{d} \vec H_c}{\mathtt{d}t} - \vec \omega_c \wedge \vec H_c = 0
\end{equation}
(Moritz and Mueller 1987) where $\vec H, \vec H_c$ are respectively the angular momentum of Venus and of the core (see also e.g. Rambaux et al 2007). The vector $\Gamma$ is the gravitational torque acting on Venus. Here, we assume that the core has a simple motion has suggested by Poincar\'e in 1910 and we neglect the core-mantle friction arising at the core-mantle boundary. Then, the rotational motion of each layers of Venus is integrated simultaneously with the gravitational torque due to the Sun acting on the triaxial figure of Venus. The orbital ephemerides DE421 (Folkner et al., 2008) are use for the purpose.

First we double-checked the good agreement between the numerical and the analytical solutions given by Eq.(\ref{eq7}) for a simplified model of one layer solid rigid Venus case. The very small differences obtained (at the level of a relative $10^{-10}$) may result from the use of different ephemerides (DE421 for the numerical approach and VSOP87 for the analytical one). Then we applied the procedure for the two layers case. As the moments of inertia $I_{c}$ of the core are not yet constrained by measurements, we used internal models (Yoder, 1995) for which $I_{c}$ is taken from $0.01$ to $0.05$, according to the size of the core. Notice that the flattening of the core is scaled to the flattening of the mantle by the assumption that the distribution of mass anomalies is the same. Table \ref{tab:core} shows the new amplitudes of $\frac{\Delta \omega}{\overline{\omega}}$ for the 4 main oscillations with arguments $2L_{s}-2\phi, 2L_{s}-2\phi+M, 2L_{s}-2\phi-M, 2\phi$ and with corresponding periods $58.37$d, $46.34$d, $78.86$d, $121.51$d. The presence of the fluid core allows a differential rotation of the mantle and of the interior of the planet. As a consequence in first approximation the mantle is decoupled from the interior and presents amplitude of libration larger than in the solid case. The amplitude of the libration increased with the size of the core. At maximum, for $I_{c}=0.05$, we obtain an increase in the oscillation of $2L_S-2\phi$ of $17\%$ corresponding to a variation of $+20.4$s. 

\begin{table}[!h]
\begin{center}
\resizebox{1.\hsize}{!}{\begin{tabular}[h]{c|cccc}
Period / 	&	58.37d  & 46.34d 	&	78.86d  & 121.51d \\
$I_{c}$  &$2L_{s}-2\phi$ &$2L_{s}-2\phi+M$&$2L_{s}-2\phi-M$ & $2\phi$\\
\hline
\hline
0.01	& 2.859979	&	0.053456	&	0.013032&		0.006328 \\
0.02	&2.950486		&0.055147		&0.013445		&0.006528	\\
0.03	&3.046908		&0.056950		&0.013884		&0.006741\\
0.04	&3.149846		&0.058874		&0.014353		&0.006968\\
0.05	&3.259981		&0.060932		&0.014855		&0.007211\\
\hline
-	&2.774866 	&	0.052244 	&	0.012726	&	0.006120
\end{tabular}
}
\end{center}
\caption{Resulting amplitude for Venus $\frac{\Delta \omega}{\overline{\omega}}$ with a fluid core (expressed in unit of $10^{-6}$). The last line is computed for a model without a fluid core.}
\label{tab:core}
\end{table}%

\begin{figure}[htbp]
\center
\resizebox{0.9\hsize}{!}{\includegraphics{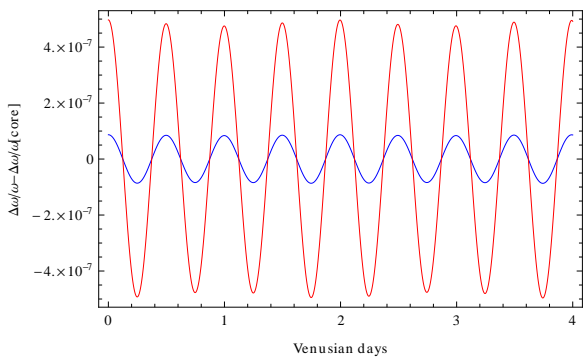}}
 \caption{$\frac{\Delta \omega}{\overline{\omega}}$ due to the core during a 4 venusian days time span with a $I_{c}=0.05$ (red curve) and a $I_{c}=0.01$ (blue curve).}
\label{fignoyau}
\end{figure}

Figure \ref{fignoyau} shows the relative variation of the speed of rotation of Venus due to the core during a 4 venusian days time span with $I_{c}=0.05$ (red curve) and $I_{c}=0.01$ (blue curve). These variations have peak to peak amplitudes respectively of $1.7\,10^{-7}$ and $9.7\,10^{-7}$. Using the definition of the $LOD(t)$ given in Section \ref{def}, the core contributions to Venus rotational speed correspond to peak to peak variations of the $LOD(t)$ between $3.6$s and $20.4$s. Although the core effects on the rotation of Venus increase with the core size, they are smaller than the solid and atmospheric effects. Indeed, when $I_{c}=0.05$ the core effects are respectively nearly two and six times smaller than the atmospheric and solid ones.

\section{Comparison and implication for observations}\label{discussion}

The variations of the rotation of Venus presented in this paper are quasi-periodic and mainly due to three kinds of effects: solid, atmospheric and core. Compared to them, the zonal potential has a negligible influence. Finally $\omega(t)$ can be written in the form:
\begin{eqnarray}\label{yy}
\omega(t)&&=\bar{\omega}+\Delta \omega\nonumber\\&&=\bar{\omega}+\Big[\sum_{i} (a_{s,i}+a_{c,i})\cos(\omega_{i}t+\rho_{i})\nonumber\\&&
+\sum_{j}a_{a,j}\cos(\omega_{j}t+\rho_{j})\Big]\bar{\omega}, 
\end{eqnarray}
where $\omega_{i}$, $\rho_{i}$ and $a_{s, i}, a_{c, i}$  (see Table \ref{table1}) are the frequencies, the phases and the corresponding amplitudes of the variation of rotation of Venus due to the solid and the core, and $\omega_{j}$, $a_{a, j}$ (see Table \ref{table2}) those due to the atmosphere. The atmospheric coefficients as well as their periods have been obtained from a fast fourier transform (FFT) where $\rho_{j}$ are the phases. Note that the power spectrum of the atmospheric variation is complex and can vary significantly from one model to another. As a consequence, only the most important frequencies are shown for the $GCM2$ model in Table~\ref{table2}.

\begin{table}[!h]
\begin{center}
\resizebox{1\hsize}{!}{\begin{tabular}[h]{c|ccc}
Period $P_{i}$& Argument &solide$\,a_{s, i}$ & Core $\,a_{c, i}$\\

$\frac{2\pi}{\omega_{i}}$&  &$\cos \omega_{i}t$&$\cos\omega_{i}t$ \\
\hline \hline\\

$58$d & $2L_{s}-2\phi$ & $2.77 \,10^{-6}$& $8.51\,10^{-8}<a_{c}<4.85\,10^{-7}$\\

$46.34$d & $2L_{s}+M-2\phi$ & $5.24 \,10^{-8}$&$1.01\,10^{-9}<a_{c}<8.5\,10^{-9}$\\

$78.86$d & $2L_{s}-M-2\phi$ & $1.27 \,10^{-8}$&$3.1\,10^{-10}<a_{c}<2.1\,10^{-9}$\\

$121.80$d & $2\phi$ & $6.12 \,10^{-9}$&$2.1\,10^{-10}<a_{c}<1.1\,10^{-9}$\\

\end{tabular}
}
\end{center}
\caption{Variation of $\frac{\Delta \omega}{\bar{\omega}}$ due to the solid and core effects.}
\label{table1}
\end{table}

\begin{table}[!h]
\begin{center}
\resizebox{0.6\hsize}{!}{\begin{tabular}[h]{c|cc}
Period $P_{j}$ &$a_{a, j} $&$\rho_{j}$\\
\hline\\
$\frac{2\Pi}{\omega_{j}}$& $\cos \omega_{j}t+\rho_{j}$\\
\hline \hline\\

$117$d &  $4.17 \,10^{-7}$&$3.11$\\
$266$d &   $3.22 \,10^{-7}$&  $1.74$\\
 $5.02$d &$1.74 \,10^{-7}$  &$2.39$\\
$6.09$d & $1.41\,10^{-7}$ & $3.14$\\
 $40.6$d&$1.28\,10^{-7}$&$0.0061$\\
$79.5$d&  $1.21\,10^{-7}$ &$1.64$\\

\end{tabular}
}
\end{center}
\caption{Variations of $\frac{\Delta \omega}{\bar{\omega}}$ due to the atmosphere effects modeled by $GCM2$.}
\label{table2}
\end{table}

Comparing the amplitudes given in Tables \ref{table1} and \ref{table2}, we see that the most important effect on the rotation rate of Venus is due to the solid potential exerted by the Sun on its rigid body. If all effects are taken into account, the variations of the $LOD(t)$ can reach $3$mn which could be observable in the future.

\begin{figure}[htbp]
\center
\resizebox{0.9\hsize}{!}{\includegraphics{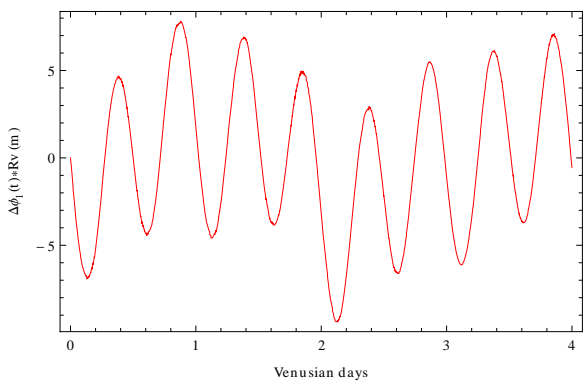}}
 \caption{Variations of $\Delta \phi_{1}\times R_{V}$ in meter during a four venusian days time span.}
\label{fig5}
\end{figure}

As most past studies used infrared imaging of the surface to measure the evolution in time of the longitude of reference points $\phi_{1}$ at Venus surface, let us consider how this variable behave in our model. As $\dot{\phi_{1}}\approx\omega (t)$ we have
\begin{equation}\label{phi}
\phi_{1}(t)=\phi_{1, 0}+ \overline{\omega}t+\int_{t_{0}}^{t} {\Delta \omega}\mathtt{d}t.
\end{equation}
Fig.\ref{fig5} shows the variations $\Delta \phi_{1}$ during four venusian days time span, caused by the combined effects of the solid, the core and the atmosphere modeled by the GCM2.  

\begin{figure}[htbp]
\center
\resizebox{0.9\hsize}{!}{\includegraphics{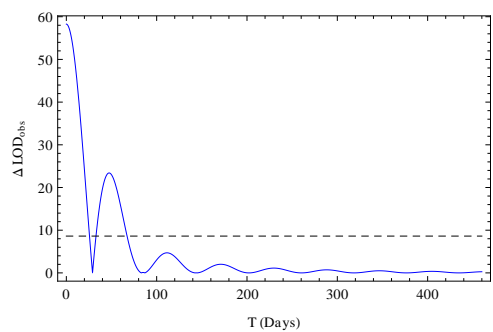}}
 \caption{Maximum error of  $\Delta LOD_{obs}(T)$ in seconds given by Eq.(\ref{erreur}). The dotted line is the uncertainty given by Magellan (Davies, 1992) on the rotation of Venus.}
\label{fig8}
\end{figure}

Like the variations of the rotation rate, these variations are periodic with a peak to peak amplitude reaching $12$ m (time differences are converted in distance at the surface). In precedent studies that measured the mean rotation rate of Venus, these variations where neglected. This implied fitting the measurements of the phase angle $\phi_{1}$ with a line of constant slope, this slope giving the mean rotation rate. Because the variations of the rotation rate discussed above are not negligible, we show in the following that this approach can yield large errors on the derived value of $\bar{\omega}$, especially for a short interval of time. Similarly to Laskar and Simon (1988) let us consider the error made on the mean rotation rate when fitting the signal given by Eq.(\ref{phi}), keeping only the most important frequency for simplicity, by a function of the type :
\begin{equation} 
\phi_{1}(t )=\phi_{1,\mathtt{obs}}+\omega_{\mathtt{obs}}\,t
\end{equation} 
over a time span $[t_{0},t_{0}+T]$.
For a least square fitting procedure, the residual is given by :
\begin{equation}
D=\int_{t_{0}}^{t_{0}+T}(\phi_{1,0}+\bar{\omega}t+A \sin \omega_{1}t-(\omega_{obs}\,t+\phi_{1,obs}))^2\mathtt{d}t
\end{equation}
where $A$ and $\omega_{1}=\frac{2\pi}{58}$rd/d correspond respectively to the larger amplitude of Eq.(\ref{phi}) and its corresponding frequency. Minimizing this residual yields
\begin{eqnarray}
\omega_{obs}&&=\bar{\omega}-\frac{6A}{T^3 \omega_{1}^2}
\Biggl[T\omega_{1}\Bigl[\cos (\omega_{1}t_{0})+\cos \omega_{1}(T + t_{0})\Bigr]\nonumber\\&&
+2 \Bigl[\sin (\omega_{1} t_{0})- \sin \omega_{1}(T + t_{0}) \Bigr]\Biggr]
\end{eqnarray}
\begin{eqnarray}
\phi_{1, obs}&&=\phi_{1,0}+\frac{2A}{T^3 \omega_{1}^2}
\Biggl[T (2 T + 3 t_{0})  \omega_{1} \cos (\omega_{1} t_{0})\nonumber\\&&+T (T + 3t_{0}) \omega_{1} \cos \omega_{1}(T + t_{0}) \nonumber\\&&+3 (T + 2 t_{0}) \Bigl[\sin (\omega_{1} t_{0})- \sin \omega_{1}(T + t_{0}) \Bigr]\Biggr]
\end{eqnarray}
We can see that $\omega_{obs}$ depends on both the time of the first observation ($t_{0}$) (i.e phase) and interval ($T$) between observations. If the phase of the effect is unknown, the max error made on the mean rotation rate is given by :
\begin{equation}\label{erreur}
\Delta LOD_{obs}=\max_{t_{0}} |\omega_{obs}(t_{0},T)-\bar{\omega}|\frac{2\pi}{\bar{\omega}^2}
\end{equation}
Fig.\ref{fig8} shows this maximum error $\Delta LOD_{obs}$ in seconds as a function of the interval $T$ between observations.

As we can see, even if the mean angular velocity is retrieved for long baseline observations, the error yielded by the linear fit can be large for short duration observations. In addition, the use of such a simple model for $\phi_{1}$ prevents any measurement of the amplitude of the variations detailed in this paper. To measure the amplitude of these variations, modeling $\omega$ and $\phi_{1}$ with Eq.(\ref{yy}) and Eq.(\ref{phi}) respectively during the data reduction is necessary.

As the most important effect is due to the torque of the Sun on the Venus rigid body with a large amplitude on a $58$d interval, it could be interesting to substract the measured signal by a fitted sinusoid of this frequency. The direct measurement of the amplitude of the sinusoid would give information on the triaxiality of Venus at $3$ to $17 \%$ of error because of the core contribution.

To disentangle atmospheric effects, a multi frequency analysis of the data will be necessary. Indeed after having determined the larger amplitude on $58$d as explained previously, it could be substracted to the analysis. Then as the atmospheric winds (described by Eq.(\ref{atmosp})) are the second most important effects on the rotation of Venus, the residuals obtained could constrain their strength. 
In parallel it should be interesting, also, to fit the signal obtained by a sinusoid on the period of $117$d because only the atmosphere acts on the rotation with this periodicity. Note that this period is also present when using an alternative atmospheric model $GCM1$ (Lebonnois, 2010a). So the corresponding amplitude could directly give indications on the atmospheric winds. Fitting the signal with additional sinusoids with other periods given in Table \ref{table2} could also increase the constraints.

At last, as the core has not the same contribution on each period presented in Table \ref{table1}, fitted the signal by Eq.(\ref{yy}) (or by Eq.(\ref{phi})) could confirm the presence of a fluid core. Of course, such a detection is possible only if the precision of the measurements is of the same order of magnitude than the core effects and if the atmospheric effects which add noise in the signal are better modelled in the future.

\begin{figure}[htbp]
\center
\resizebox{0.9\hsize}{!}{\includegraphics{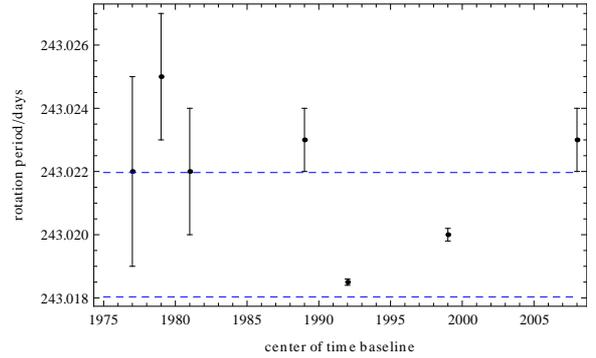}}
 \caption{Values of the period of rotation of Venus and their error bars given since $1975$. The dotted lines represent variations of $0.00197$d around the mean values of $243.020$d caused by the effects presented in this paper.}
\label{fig9}
\end{figure}

Many values of the mean rotation of Venus have been estimated since 1975. Fig.\ref{fig9} shows all these values as well as their error bars. The latter value of $243.023\pm0.001$d has been recently estimated from Venus express VIRTIS images (Mueller et al., submitted paper 2010). The dotted lines represent variations of $0.00197$d around mean value of $243.020\pm0.0002$d (Konopliv et al., 1999) caused by solid, atmospheric ($GCM2$), and core ($I_{c}=0.05$) effects presented in this paper. Note that the value of Davies et al. (1992) set to $243.0185\pm0.0001$d has been recommended by the IAU (Seidelmann et al., 2002).

As we can see from Fig.\ref{fig9}:
\begin{itemize}
\item
the variations of the period of rotation ($0.00197$d) presented in this paper are consistent with most of the mean rotation periods measured so far.
\item
if the true mean value of the period of rotation of Venus is close to the IAU value, the variations cannot explain the most recent value obtained by VIRTIS (Mueller et al., submitted paper 2010). Indeed, the difference between them of $7$mn implies larger variations.
\end{itemize}

The different values of the rotation of Venus since 1975 could be explained by the variations of $0.00197$d due to the solid, atmospheric and core effects. Despite the fact that the value of Davies et al. (1992) has been recommended by the IAU because of its small published error bars ($\pm 0.0001$), its large difference with the recent VIRTIS value ($7$mn) and the other measurements, is not in agreement with the variations presented in this paper. It seems difficult to explain these large variations, which would imply an increase of more than $50\%$ of the effects discussed here, with the current models. So it would be interesting to compare in detail the different rotation rate measurement methods and to determine with a better accuracy the mean value of the rotation of Venus.

\section{Conclusion}\label{conclusion}

The purpose of this paper was to detemine and to compare the major physical processes influencing the angular speed of rotation of Venus. Applying different theories already used for the Earth, the variation of the rotation rate as well as of the $LOD$ was evaluated. 

Applying the theory derived from Kinoshita (1977), the effect of the solid potential exerted by the Sun on a rigid Venus was computed. Considering in the first step that the orbit of Venus is circular, we found that the variations of the rotation rate have a large amplitude of $2.77\,10^{-6}$ with argument $2L_{s}-2\phi$ ($58$d). Taking into account the eccentricity of the orbit adds periodic variations with a $10^{-8}$ amplitude. On the opposite of the Earth, as Venus has a very slow rotation, the solid potential has a leading influence on the rotation rate which corresponds to peak to peak variations of the $LOD$ of $120$s

Considering Venus as an elastic body, we then evaluated the impact of the zonal tidal potential of the Sun. These variations correspond to peak to peak variations on the $LOD$ of $0.014$s which are very small with respect to the contributions of the solid effect.

Then we computed the effects of both the core and the atmosphere (modeled by two different simulations of the LMD global circulation model). According to our computation the atmospheric and core contributions to Venus rotational speed correspond to peak to peak variations of the $LOD$ of $25-50$s and $3.5-20.4$s respectively at different periods. Despite its thickness, the impact of the venusian atmosphere modeled by our most realistic simulation on the rotation is $2.4$ times smaller than the contribution of the solid torque exerted by the Sun. The variations of the $LOD$ due to the core, which increase with its size, are still smaller.

At last we have shown that the variations of $\omega$ and $R_{\mathtt{V}}\phi_{1}$ which reach $3$mn and $12$m respectively, need to be taken into account in the reduction of the observations. Ignoring these variations could lead to an incorrect estimation of $\bar{\omega}$. 
With the steadily increasing precision of the measurements, carrying a frequency analysis of the data modeled by either Eqs.(\ref{yy}) or (\ref{phi}) will hopefully enable to put physical constraints on the physical properties of Venus (triaxiality, atmosphere, core).

To conclude, the variations shown in this paper, would explain different values of the mean rotation of Venus given since 1975. The difference of $7$ mn between the IAU value ($243.0185$, Davies et al., 1992) and the VIRTIS one ($243.023\pm0.001$, Mueller et al., submitted paper, 2010) implies larger variations, not found here, probably due to systematic errors in the measurements of the rotation. In view of these new results and the recent study of Mueller et al. (submitted paper, 2010), it would be interesting to compare the different  methods of measurement of the rotation of Venus, to identify the error sources with a better accuracy and to revise the value of the mean rotation.

\begin{acknowledgements}
L.C wishes to thanks Pierre Drossart, Thomas Widemann, Jean-Luc Margot and Jeremy Leconte for discussion on Venus.
\end{acknowledgements}

\appendix
\section{Influence of the angle $J$ on the rotation rate}

In Section \ref{solide}, we assumed that the angle $J$ between the angular momentum axis and the figure axis of Venus can be neglected.  Here we reject this hypothesis of coincidence of the poles and we evaluate the impact of $J$ on the rotation. According to Eq.\ref{eq1} the variations of  $\Delta \omega$ is given by :
\begin{equation}\label{AO}
\frac{\mathtt{d}}{\mathtt{d}t}(C\Delta \omega)= -\frac{\partial U}{\partial g}. 
\end{equation}
where 
\begin{eqnarray}\label{A1}
U&&=\frac{\mathcal{G} M'}{r^3}\Big[[\frac{2C-A-B}{2}]P_{2}(\sin \delta )\nonumber\\&&+[\frac{A-B}{4}]P_2^{2} (\sin \delta) \cos 2\alpha\Big],
\end{eqnarray}
As it was done in Section \ref{solide}, we express $P_{2}(\sin \delta )$ and $P_2^{2} (\sin \delta) \cos 2\alpha$ as functions of the longitude $\lambda$ and the latitude $\beta$ of the Sun with the transformation described by Kinoshita (1977) without supposing $J=0$ such as:
\begin{eqnarray}\label{A2}
P_{2}(\sin \delta)&\nonumber& = \frac{1}{2}(3\cos^2 J-1)\Bigg[\frac{1}{2}(3\cos^2 I-1)P_{2}(\sin \beta)\nonumber \\&&
-\frac{1}{2} \sin 2I \sin(\lambda-h) P_{1}^2(\sin \beta) \nonumber \\&&-\frac{1}{4} \sin 2I  P_{2}^1(\sin \beta)\cos 2(\lambda - h)\Bigg]\nonumber \\&&+\sin 2J \Bigg[-\frac{3}{4}\sin 2I P_{2}(\sin \beta)\cos g \nonumber \\ &&-\sum_{\epsilon =\pm 1}\frac{1}{4}(1+\epsilon \cos I)(-1+2\epsilon \cos I) \nonumber \\&& P_{2}^1(\sin \beta)\sin(\lambda-h-\epsilon g) \nonumber \\&& -\sum_{\epsilon =\pm 1}\frac{1}{8}\epsilon \sin I(1+\epsilon \cos I)\nonumber \\&& P_{2}^2(\sin \beta)\cos(2\lambda-2h-\epsilon g)\Bigg]\nonumber\\&&
+\sin^2 J \nonumber \\&& \Bigg[\frac{3}{4}\sin^2I P_{2}(\sin \beta)\cos 2g+\frac{1}{4}\sum_{\epsilon =\pm 1}\epsilon \sin I  \nonumber \\&& (1+\epsilon \cos I)P_{2}^1(\sin \beta)\sin(\lambda-h-2\epsilon g)-\frac{1}{16}\nonumber \\&& \sum_{\epsilon =\pm 1}(1+\epsilon \cos I)^2P_{2}^2(\sin \beta)\cos 2(\lambda-h-\epsilon g)\Bigg]\nonumber \\
\end{eqnarray}
and
\begin{eqnarray}
&P_{2}^2(\sin \delta)\nonumber&\cos 2\alpha =3\sin^2 J\Bigg[-\frac{1}{2}(3\cos^2 I-1)P_{2}(\sin \beta)\nonumber \\&& \cos 2l+\frac{1}{4}\sum_{\epsilon =\pm 1}\sin 2IP_{2}^1(\sin \beta)\sin(\lambda-h-2\epsilon l) \nonumber \\&& +\frac{1}{8}\sin^2 IP_{2}^2(\sin \beta)\cos 2(\lambda-h-\epsilon l)\Bigg]\nonumber \\ &&+\sum_{\rho =\pm 1}\rho \sin J (1+\rho \cos J)\nonumber \\&& \Bigg[-\frac{3}{2}\sin 2IP_{2}(\sin \beta) \cos (2\rho l+g)\nonumber \\&&-\sum_{\epsilon =\pm 1}\frac{1}{2}(1+\epsilon \cos I)(-1+2\epsilon \cos I)\nonumber \\&&  P_{2}^1(\sin \beta)\sin(\lambda-h-2\rho \epsilon l-\epsilon g)\nonumber \\&& -\sum_{\epsilon =\pm 1}\frac{1}{4}\epsilon \sin I(1+\epsilon \cos I) \nonumber \\&&  P_{2}^2(\sin \beta)\cos(2\lambda-2h-2\rho\epsilon l-\epsilon g)\Bigg]\nonumber
\end{eqnarray}
\begin{eqnarray}
&&+\sum_{\rho =\pm 1}\frac{1}{4}(1+\rho \cos J)^2\Bigg[-3\sin^2 IP_{2}(\sin\beta)\nonumber \\&&\cos(2l+2\rho g)\sum_{\epsilon =\pm 1}\epsilon \sin I(1+\epsilon \cos I)\nonumber \\&& P_{2}^1(\sin\beta) \sin(\lambda -h-2 \rho \epsilon l-2\epsilon g)\nonumber \\&&+\sum_{\epsilon =\pm 1}\frac{1}{4}(1+\epsilon \cos I)^2 \nonumber \\&&  P_{2}^2(\sin\beta)\cos 2(\lambda-h-\rho\epsilon l- \epsilon g)\Bigg].\nonumber \\
\end{eqnarray}
Assuming that $\beta=0$ and removing the components which do not depend on the variable $g$ this yields:
\begin{eqnarray}\label{A3}
\frac{\mathtt{d}(C\Delta \omega)}{\mathtt{d}t}&&=-\frac{\partial }{\partial g}\frac{\mathcal{G} M'}{a^3} \Big(\frac{a}{r}\Big)^3
\Bigg[\frac{2C-A-B}{2}\nonumber\\&&
\Bigg(\sin 2J \bigg[\frac{3}{4}\sin 2I \frac{1}{2}\cos g \nonumber \\ && -\sum_{\epsilon =\pm 1}\frac{1}{8}\epsilon \sin I(1+\epsilon \cos I)\nonumber \\&& 3\cos(2\lambda-2h-\epsilon g)\bigg]\nonumber\\&& 
+\sin^2 J \nonumber \\&& \bigg[-\frac{3}{4}\sin^2I \frac{1}{2}\cos 2g-\frac{1}{16}\nonumber \\&& \sum_{\epsilon =\pm 1}(1+\epsilon \cos I)^2 3\cos 2(\lambda-h-\epsilon g)\bigg]\Bigg)\nonumber\\&& 
+\frac{A-B}{4}\Bigg(\sum_{\rho =\pm 1}\rho \sin J (1+\rho \cos J)\nonumber \\&& \Big[\frac{3}{2}\sin 2I  \frac{1}{2} \cos (2\rho l+g)\nonumber \\&& -\sum_{\epsilon =\pm 1}\frac{1}{4}\epsilon \sin I(1+\epsilon \cos I) \nonumber \\&&  3 \cos(2\lambda-2h-2\rho\epsilon l-\epsilon g)\Big]\nonumber\\&&
+\sum_{\rho =\pm 1}\frac{1}{4}(1+\rho \cos J)^2\Big[3\sin^2 I  \frac{1}{2} \nonumber \\&&\cos(2l+2\rho g)+\sum_{\epsilon =\pm 1}\frac{1}{4}(1+\epsilon \cos I)^2 \nonumber \\&&  3\cos 2(\lambda-h-\rho\epsilon l- \epsilon g)\Big]\Bigg)\Bigg],
\end{eqnarray}
where $\frac{\mathcal{G}  M'}{r^3}$ has been replaced by  $n^2 (\frac{a}{r})^3$. 

\begin{figure}[htbp]
\center
\resizebox{0.9\hsize}{!}{\includegraphics{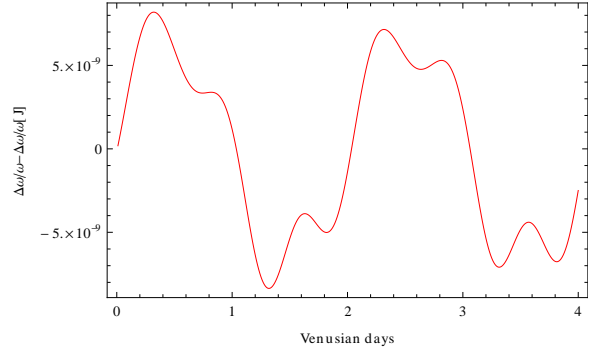}}
 \caption{Influence of the angle $J=0.5^{\circ}$ on the variation of the speed of rotation of Venus.}
\label{fig11}
\end{figure}

The variations of the speed of rotation of Venus due to the solid torque exerted by the Sun are obtained by developing Eq.(\ref{A3}) as functions of time through the variables $M$ and $L_{s}$ taking the eccentricity as a small parameter (Kinoshita, 1977). Fig.\ref{fig11} shows the residuals after substraction of $\frac{\Delta \omega}{\omega}$ obtained by the Eq.(\ref{A3}) when the angle $J=$ is taken into account and $e=0$ with respect to the variations given by Eq.(\ref{eq7}) in Section \ref{solide}. We arbitrarily take the angle $J=0.5^{\circ}$.

The influence of the angle $J=0.5^{\circ}$ on the rotation with amplitudes of  $\approx 10^{-9}$ is smaller than the leading coefficients seen in Eq.(\ref{eq7}) by three orders of magnitude and than the influence of the eccentricity of Venus by one order of magnitude. As J is probably much smaller than $0.5^{\circ}$ as it is the case for the Earth ($1"$), its influence on the rotation can be neglected with respect to the other effects taken into account in this paper.

\end{document}